# Tuning capillary flow in porous media with hierarchical structures

Si Suo[1], and Yixiang Gan[1,2] *

[1] School of Civil Engineering, The University of Sydney, NSW 2006, Australia

[2] Sydney Nano, The University of Sydney, NSW 2006, Australia

* Corresponding author: yixiang.gan@sydney.edu.au (Y. Gan)

**Abstract:**

Immiscible fluid-fluid displacement in porous media is of great importance in many engineering applications, such as enhanced oil recovery, agricultural irrigation, and geologic $CO_2$ storage. Fingering phenomena, induced by the interface instability, are commonly encountered during displacement processes and somehow detrimental since such hydrodynamic instabilities can significantly reduce displacement efficiency. In this study, we report a possible adjustment in pore geometry which aims to suppress the capillary fingering in porous media with hierarchical structures. Through pore-scale simulations and theoretical analysis, we demonstrate and quantify combined effects of wettability and hierarchical geometry on displacement patterns, showing a transition from fingering to compact mode. Our results suggest that with a higher porosity of the $2^{nd}$-order porous structure, the displacement can keep compact across a wider range of wettability conditions. Combined with our previous work on viscous fingering in such media, we can provide a complete insight into the fluid-fluid displacement control in hierarchical porous media, across a wide range of capillary number from capillary- to viscous-dominated modes. The conclusions of this work can benefit the design of microfluidic devices, as well as tailoring porous media for better fluid displacement efficiency at the field scale.

**Graphic abstract:**

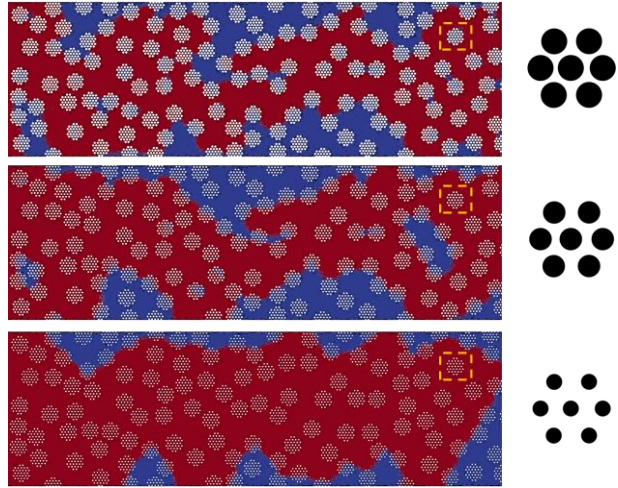

# 1. Introduction

Immiscible fluid-fluid displacement in porous media has been widely observed in various settings, such as $CO_2$ sequestration [1-3], liquid drainage in polymer-based fuel cells [4], enhanced oil/gas recovery [5, 6]. With the displacement proceeding, fingering phenomena induced by the interface instability may occur leading to a significant reduction of displacement efficiency and a ramified fluid morphology compared with a compact displacement. Understanding what controls the displacement pattern and furthermore to what extent the fingering can be suppressed are beneficial to all these applications and also essential for estimating the post-displacement status. For instance, the production of oil is directly related to the displacement efficiency and the relative permeability of the residual oil impacts the secondary recovery strategy [7, 8].

The interface instability is a result of combined effects of viscosity and capillarity which is suggested to be quantified using capillary number, $Ca$, and viscosity ratio, $M$. Various combination leads to the displacement pattern transition among viscous fingering, capillary fingering and compact displacement, as depicted in many phase diagrams [9-13]. Some studies also indicate that altering wettability shows certain potential to suppress fingering phenomena since the capillary effect is directly related to the contact angle, and specifically the compact mode covers a wider range of regimes when the solid surface is more hydrophilic to the invading fluid [14-16]. However, a recent study by Zhao et al. [17] points that for the strong imbibition case, i.e., the surface is of extreme affinity to the invading fluid, displacement instability may be triggered due to the corner flows. Other possible ways, including using non-Newtonian fluids [18, 19] or additional electrical field [20], can weaken the interface instability for some specific situations.

The geometry of the porous matrix also plays a significant role on the multiphase flow, especially on the pressure distribution and capillary effects [21-23]. For an ordered porous medium, the pore size gradient along the flow direction has been proven an effective approach of suppressing viscous fingering [24] as well as capillary fingering [25]. For a disordered porous medium, recent studies [26, 27] suggest that the displacement tends to be compact with reducing disorder. However, it is still unclear for a given porous media with highly disordered topology that how the displacement pattern is controlled by geometry and to what extent the instability can be mitigated.

Capillary fingering naturally occurs during a capillary force dominated flow characterized typically by $M \geq 1$ and $Ca \ll 1$. Different from viscous fingering, the invading morphology of capillary fingering highly depends on the pore geometry [28] especially for the imbibition situation (i.e., the wetting phase invading the nonwetting one) since the interface advances spontaneously along solid surfaces [29]. In addition to the different dominating mechanisms in resulting flow patterns in both viscous and capillary regimes, the challenge in modelling capillary fingering resides in the dramatically increased computational cost, due to the much slower flow conditions ($Ca \ll 1$). In our recent work [30], we have demonstrated that the viscous fingering can be suppressed in a homogeneous porous medium by adopting the hierarchical porous structure. Hence, in this work, we investigate the displacement patterns through pore-scale simulations in a capillary-dominated situation, as the complementary piece in understanding the complete flow behaviour, with a special focus on adjusting the hierarchical geometry as a possible way to control capillary flows in disordered porous media. The transition from fingering mode to compact mode is demonstrated as in a phase diagram showing the combination of wettability and hierarchical geometry. By analysing the evolution of related indices during the displacement, we discover the mechanism dominating the fingering suppression and further quantitatively characterise it using a dimensionless number, which incorporates the capillary suction as a driving force instead of the external pressure in a viscous-dominated situation. The current work provides a complete picture on tuning the fluid displacement mode in hierarchical porous media, by drastically extending the capillary number regime by several orders of magnitudes.

## 2. Numerical method

As proved by many studies [31, 32], the volume of fluid (VOF) method is a well-developed and practical numerical solution for multiphase flow problems at pore scale. Here, we briefly introduces the governing equations for impressible two-phase flows, i.e., the continuity equation (1), phase fraction equation (2), and momentum equation (3) as

$$\nabla \cdot \boldsymbol{u} = 0, \tag{1}$$

$$\frac{\partial \vartheta}{\partial t} + \nabla \cdot (\boldsymbol{u} \cdot \vartheta) + \nabla \cdot [\boldsymbol{u}_r \cdot \phi \cdot (1 - \vartheta)] = 0, \tag{2}$$

$$\frac{\partial (\rho \cdot \boldsymbol{u})}{\partial t} + \nabla \cdot (\rho \cdot \boldsymbol{u} \cdot \boldsymbol{u}) - \nabla \cdot (\mu \cdot \nabla \boldsymbol{u}) - (\nabla \boldsymbol{u}) \cdot \nabla \mu = -\nabla p + \gamma \cdot \kappa \cdot \nabla \phi, \tag{3}$$

where $\phi$ is the phase fraction of two fluids, $\boldsymbol{u}$ is the weighted average of velocity field shared by two fluids, i.e., $\boldsymbol{u} = \vartheta \cdot \boldsymbol{u}_{f1} + (1 - \vartheta) \cdot \boldsymbol{u}_{f2}$, and $\boldsymbol{u}_r$ is the relative velocity, i.e., $\boldsymbol{u}_r = \boldsymbol{u}_{f1} - \boldsymbol{u}_{f2}$, $\rho$ and $\mu$ represent the weighted average of density and viscosity, respectively, i.e., $\rho = \vartheta \cdot \rho_{f1} + (1 - \vartheta) \cdot \rho_{f2}$ and $\mu = \vartheta \cdot \mu_{f1} + (1 - \vartheta) \cdot \mu_{f2}$, $p$ is the pressure, $\gamma$ is the surface tension, and $\kappa$ is the mean curvature of the interface between two fluids. For more details regarding the treatment of inlet-outlet boundaries and wetting condition of solid surfaces, one can refer to [33, 34].

Our study is on a 2D domain initially saturated with the defending phase and then displaced from the left, see Figure 1. The left side is a uniformly inlet boundary with a fixed flow rate while the outlet boundary on the right side is set with a total pressure value $p_{out} = 0$ Pa; the top and bottom sides are no-slip walls with contact angle of 90°.

Since we focus on the capillary-dominated displacement processes, it is assumed that the invading and defending fluids have the same viscosity $\eta = 1$ mPa·s, density $\rho = 1 \times 10^3$ kg/m³ and surface tension between two fluids $\gamma = 28.2$ mN/m; meanwhile the inlet flow rate is limited as a very small value, i.e., $v_{in} = 1 \times 10^{-3}$ mm/s, so that the viscosity ratio $M = 1$ and capillary number $Ca = 3.55 \times 10^{-8}$. Note that modelling the capillary-dominate regime is quite computationally intensive, due to the extreme slow flow rate and extensive simulation time. According to the phase diagram of fluid-fluid displacement patterns [13], such combination of $M$ and $Ca$ lies within the capillary fingering regime. The stability of a fluid-fluid displacement also depends on the topology of solid obstacles as suggested in Ref [26], and specifically the displacement process tend to be instable with a larger disorder index $I_v$. Thus, through the Mont Carlo iteration, we here generate a the domain of interest contains randomly-distributed round obstacles with radius $R_1 = 0.8$ mm with $I_v = 0.052$, as shown in Figure 1, which can guarantee the capillary fingering to occur during the weak imbibition (contact angle $\theta > 30°$). To shed the light on the effect of hierarchical structure on displacement patterns, each obstacle has identical homogeneous porous structure with the 2nd-order throat size ($d_2$) equalling 0.06, 0.112, 0.144 and 0.160 mm, and corresponding geometry parameters, including particle radius ($R_2$), porosity ($\phi_2$), and permeability ($k_2$) are listed in Table 1. Moreover, the simulations are performed with a group of contact angle $\theta$, i.e., ranging from 30° to 90° to cover a wide range of wettability conditions. The interval of contact angle is later refined in specific regions to illustrate the transition of displacement modes, in Section 3.

Table 1 Geometry parameters for the 2nd-order pore structures, with $\phi_1$=0.64 and $R_1$=0.8 mm.

| $d_2$ (mm) | $R_2$ (mm) | $\phi_2$ | $k_2$ (mm$^2$) |
| --- | --- | --- | --- |
| 0.060 | 0.105 | 0.47 | 2.34×10$^{-4}$ |
| 0.112 | 0.079 | 0.70 | 1.27×10$^{-3}$ |
| 0.144 | 0.0626 | 0.81 | 2.62×10$^{-3}$ |
| 0.160 | 0.0556 | 0.85 | 3.60×10$^{-3}$ |

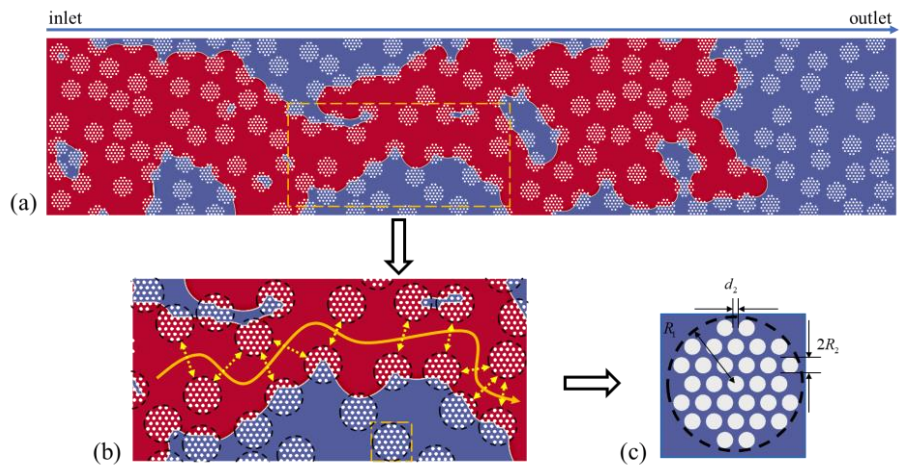

Figure 1. (a) The numerical model with two-level hierarchical porous media, i.e., (b) 1st-order disordered and (c) 2nd-order homogeneous porous structures.

## 3. Results and discussion

Generally, a fluid-fluid displacement process can be divided into pre- and post-breakthrough stages, i.e., split by the moment when the invading front reaches the outlet. The global channels for the invading phase are built at the pre-breakthrough stage and remains mostly at the post-breakthrough stage; only the local fluid distribution may be rearranged due to the unbalance of capillary pressures and such capillary rearrangement can last for a relative long time during the post-breakthrough stage [35]. The contrast between the durations of these two time periods in viscous fingering can be notable, whilst in capillary fingering, the global and local balance can be acquired in similar time scales. Furthermore, considering the major flow patterns can be characterized at the early evolution, we only investigate the displacement processes before the breakthrough in this study.

### 3.1 Phase diagram

Capillary-dominated displacement is generally regarded as a percolation-like process [29], i.e., the slow invasion advances pore by pore and prefers the pathway with the lowest entry capillary pressure. The geometric disorder leads to nonuniformly distributed capillary pressure and finally causes phase trapping as well as fingering. However, in hierarchical porous media, such macroscopic pathway-bias may be balanced due to $2^{nd}$-order porous structures. Specifically, under the same flow condition, the displacement pattern may transit from the fingering to compact mode by adjusting the $2^{nd}$-order geometry. The effects of hierarchical geometry on a capillary-dominated displacement under various wetting conditions from strong to neutral imbibition are summarized as a phase diagram in Figure 2. The displacement pattern can transit from the compact to fingering mode with increasing contact angle $\theta$ [14], which agrees with the observation for each geometry as shown in Figure 2. However, with a fixed wetting condition, cases with different $2^{nd}$-order porous structures may present evidently distinct displacement processes, especially within the regime of weak and neutral imbibition ($\theta = 60°\sim 90°$). In another word, the mode transition for each geometry shows different sensitivity to the wetting condition though the crossover boundary mostly lies within weak imbibition regime ($45° < \theta < 90°$), and specifically for dense packing cases ($\phi_2 < 0.8$) at the $2^{nd}$-order porous structure, the crossover boundary is at around $\theta \approx 53°$ while for loose packing cases ($\phi_2 > 0.8$) the critical contact angle $\theta > 60°$, i.e., the fingering mode can be suppressed within a larger range of weak imbibition regime for the looser packing cases.

Correspondingly, even though under the same wetting conditions where the fingering mode occurs, the displacement process demonstrates relatively higher stability in cases with loosely packed obstacles. Since the capillary number in this simulation is small enough so that the capillary effects govern the whole displacement, and thus the dynamics of the fluid-fluid interface is mostly driven by the capillary pressure which is determined by the pore geometry and contact angle. Before spontaneously infiltrating the pore space, the external pressure should be larger than the entry capillary pressure at the throat. For loose packing cases, the capillary effect is weaker, i.e., the entry capillary pressure is lower while the spontaneous driving pressure is also lower, than that in dense packing cases under a fixed wetting condition. The competition between the entry capillary resistance and spontaneous infiltrating leads to the transition between the above two displacement modes. Notably, for the strong imbibition regime ($\theta \leq 30°$), although the compact displacement occurs in all cases, there exist certain noticeable difference among them, especially in the case with $\phi_2 = 0.47$ and $\theta = 30°$, as shown in Figure 2(c), more defending fluid is trapped within 1st pore space along the top and bottom walls during the displacement while almost a perfect compact displacement occurs in other cases under the same wetting condition. Since the negative capillary pressure is quite larger along the narrow channels in the strong imbibition regime, the interface in the case with $\phi_2 = 0.47$ moves so much faster through the 2nd-order porous structure than along the walls or through the 1st-order pore space that defending phase is left behind and trapped finally. This phenomena is similar to the corner flow reported in Refs [15, 17], i.e., the invading fluid advances preferentially along narrow corners due to the intensive capillary effects resulting in local instability.

In summary, during a capillary-dominated displacement process, the strong capillary effects may lead to instability, and specifically the global fingering pattern occurs in the weak and neutral imbibition regime while the local instability like trapping of defending phase arises in the strong imbibition regime. In particular, for hierarchical porous media, the transition of displacement modes can be controlled by adjusting the 2$^{nd}$-order geometry, and the fingering mode can be suppressed within a large range of wetting conditions for cases with loosely packed obstacles.

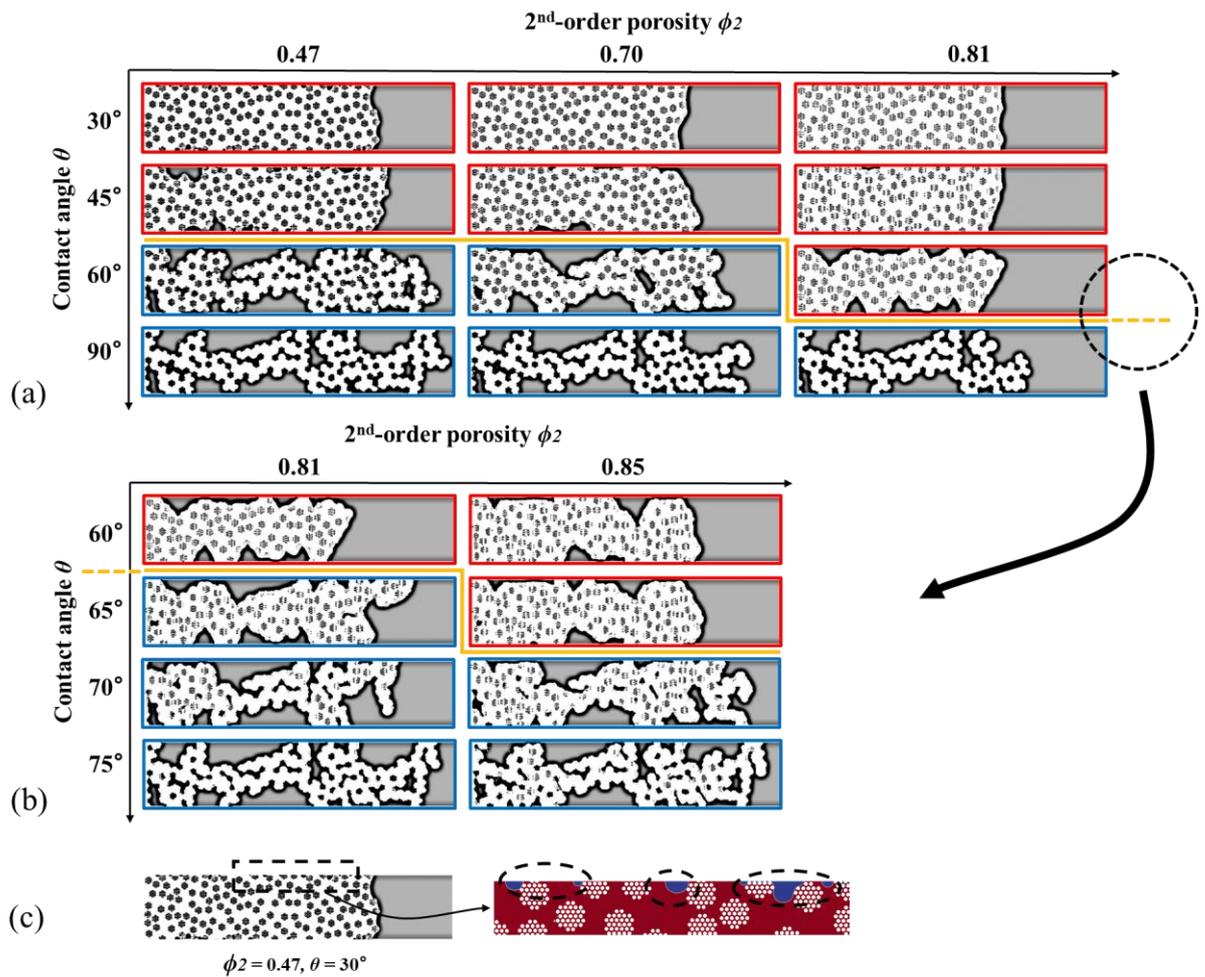

Figure 2. Phase diagrams (a-b) of displacement mode transition between the compact mode □ and the fingering mode □ in hierarchical porous media with various 2$^{nd}$-order geometry designs and different wetting conditions, and the line ⌐ indicates the position of the crossover boundary. (c) shows a zoom-in view of trapped defending fluid in case with $\phi_2 = 0.47$ and $\theta = 30°$.

## 3.2 Temporal evolutions

To quantitatively characterize the fluid-fluid displacement processes in hierarchical porous media and demonstrate the different displacement modes, a set of indexes are extracted from each time frame and demonstrated in Figure 3.

(I) The relative fluid-fluid interface length ($L_i$), i.e., the total interface length normalized by the domain width, is a good indicator to distinguish two displacement modes. Specifically, $L_i$ should be around unity for compact displacement since the invading front is almost parallel to the short side of the domain while it becomes much larger once dendrites are formed and grow. On the other hand, $L_i$ can also reflect the local instability since it is much sensitive to the fluid trapping, e.g., as shown in Figure 3(a), for the compact displacement, $L_i$ in the case with $\phi_2 = 0.47$ increases faster with respect to the injecting time than that in other cases because of the trapped defending ganglia marked in Figure 2(c). Considering that $L_i$ presents a linear relationship with the injecting time for each case, the growth rate $k_{int} = \mathrm{d}L_i/\mathrm{d}t$ is extracted here to characterize the process.

(II) Fractal dimension $D_f$, measured by box counting algorithm [36], can estimate the overall compactness of invading fluid distribution, i.e., that $D_f$ is close to 2 suggests the domain is being filled with invading phase uniformly; on the contrary, the fluid trapping or dendrite growth may occur with $D_f$ of smaller values. For each case with the displacement evolving, $D_f$ tends to reach a constant level $\widetilde{D}_f$ after 10,000 s, as shown in Figure 3(b). The crossover boundary marked in Figure 2(a-b) corresponds to a stable level of $\widetilde{D}_f \approx 1.95$.

(III) Evolution of degrees of saturation for the invading fluid in the $2^{\mathrm{nd}}$-order ($S_2$) and $1^{\mathrm{st}}$-order ($S_1$) pore space describes to what extend the invading fluid infiltrate the $2^{\mathrm{nd}}$-order pore space when the fluid advances through the main channels in the $1^{st}$-order porous structure. Considering the secondary channels in the wetted $2^{nd}$-order pore space can connect the neighbouring main channels in the $1^{st}$-order pore space, the $2^{nd}$-order saturation, $S_2$, in effect contributes to the relative permeability of the invading fluid and furthermore with higher values of $S_2$, the disorder-induced nonuniform distribution of capillary pressure can be mitigated and the instability be suppressed as a result. As shown in Figure 3(c), $S_2$ is almost linearly proportional to $S_1$ for each case, the slope $k_{sat}$ is used here to measure the involvement of $2^{nd}$-order porous structures during the displacement process. The simulation results suggest that when $k_{sat}$ is larger than 0.9, the displacement tend to be compact while the porous obstacle behaves like solid one when $k_{sat}$ is smaller than 0.4 so that the fingering may occur.

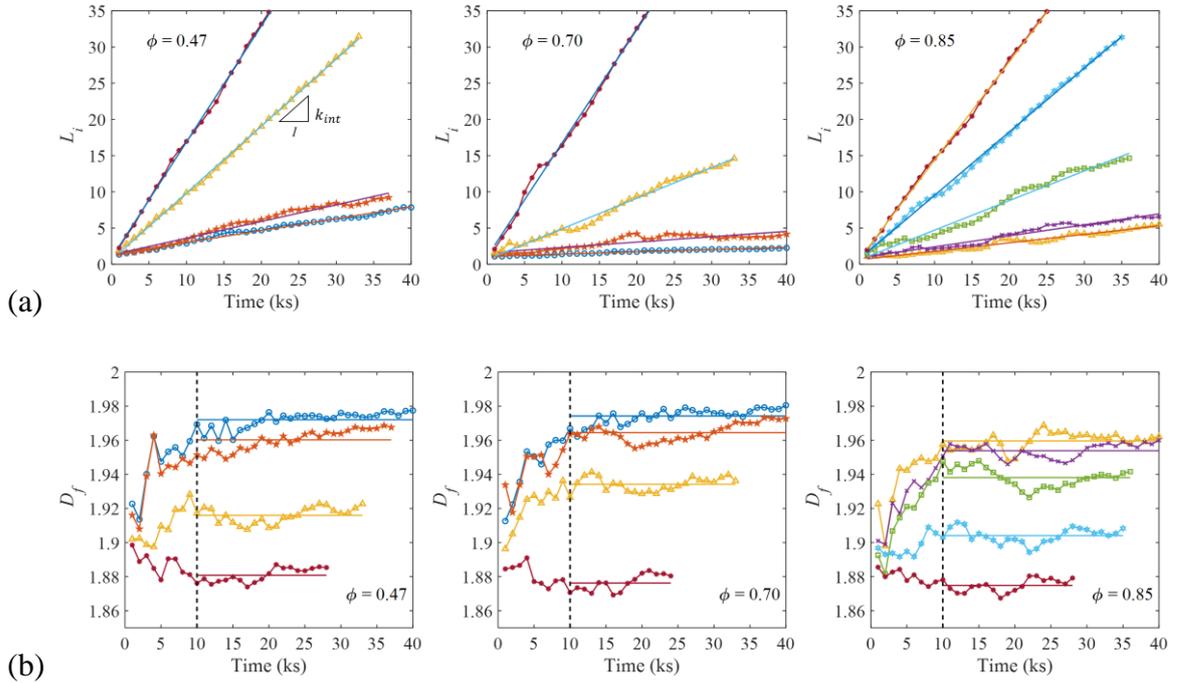

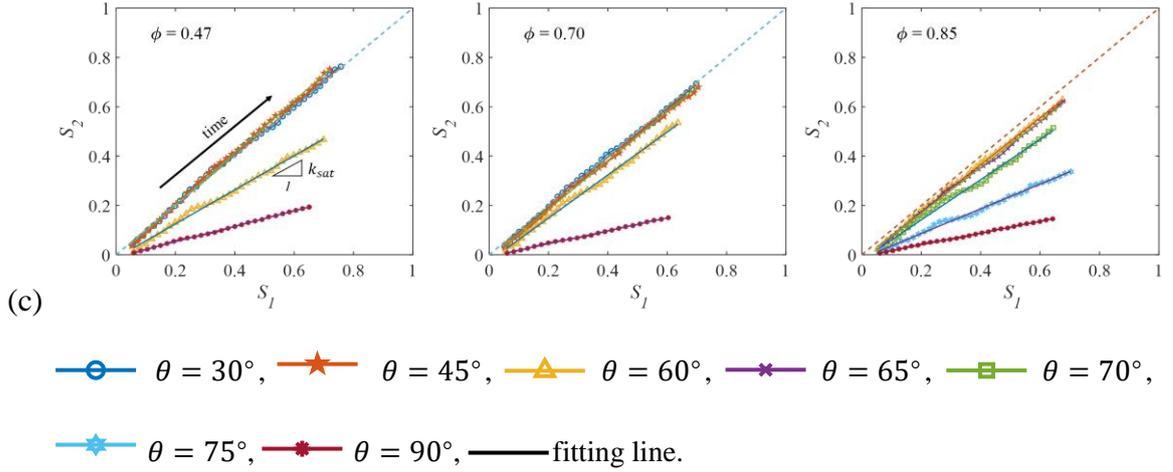

(c)

— ⊙ — $\theta = 30°$, — ★ — $\theta = 45°$, — △ — $\theta = 60°$, — × — $\theta = 65°$, — ▢ — $\theta = 70°$, — ✦ — $\theta = 75°$, — ✱ — $\theta = 90°$, —— fitting line.

Figure 3. The evolutions of displacement indexes: (a) the relative fluid-fluid interface length $L_i$ vs. injecting time; (b) the fractal dimension $D_f$ vs. injecting time and the dash line (– – –) indicates that the fractal dimension almost keeps a steady level after the marked moment; (c) Saturation in 2nd-order pore space $S_2$ vs. in 1st-order pore space $S_1$ and the dash line (– – –) indicates $S_2 = S_1$.

## 3.3 Dimensionless analysis

Based on the observation of Figure 2 and analysis on characterization of the displacement patterns, the mode transition is determined by two factors: one is the interaction time which connects the 1st-order and 2nd-order flows in hierarchical porous media; the other is the capillary competition between entry resistance and spontaneous driving force.

**(1) Interaction time**

The mechanism for fingering suppressing is that the infiltrated secondary channels in 2nd-order pore space enhances the global permeability of the invading phase by connecting the main channels in 1st-order pore space resulting in the mitigation of distribution nonuniformity of capillary pressure. So, the displacement pattern is controlled by how much invading fluid can infiltrate the 2nd-order pore space when the front advances through the 1st-order channels. To exactly reflect such effect, we use a characterized time scale ratio $R_T$ of similar expression with our previous work [30], i.e., the time ($T_{1st}$) for invading front advancing a certain characteristic distance (e.g., $R_1$) in 1st-order pore space and that ($T_{2nd}$) for capillary-driven infiltration in 2nd-order pore space as

$$T_{1st} = \frac{R_1}{v_{in}}, \tag{4}$$

$$T_{2nd} = \frac{\eta \phi_2 R_1}{2 k_2 P_c^*}, \tag{5}$$

where $P_c^*$ is the characteristic capillary pressure estimated by

$$P_c^* = \frac{\int_{-\pi/2}^{\pi/2} P_c(\alpha) d\alpha}{\pi}, \text{ and} \tag{6}$$

$$P_c = \frac{\gamma}{d_2} \frac{\cos(\theta - \alpha)}{1 + 2 R_2 / d_2 (1 - \cos(\alpha))}. \tag{7}$$

The calculation model for capillary pressure as a function of filling angle $\alpha$ in Eq. (7) is the interface advances through two cylinders with radius $R_2$ and gap distance $d_2$ and more details can be referred to Ref [30]. Finally, the time ratio, $R_T$, can be expressed as

$$R_T = \frac{T_{1st}}{T_{2nd}} = \frac{2 k_2 P_c^*}{\phi_2 v_{in} \eta R_1}. \tag{8}$$

According to Eq. (8), $R_T$ describes the interaction between two characteristic scales by combining the forced flow in 1st-order porous structure and spontaneous flow in 2nd-order pore space. With a larger $R_T$, more secondary channels are expected to be filled with the invading fluid spontaneously leading to a more compact displacement. However, the capillary effect in Eq. (8) is just roughly estimated. By considering that the entry capillary pressure also controls the spontaneous infiltration of the invading fluid, $R_T$ should be further modified.

**(2) Capillary competition**

When the invading fluid enter a pore body through a narrow throat, the negative capillary pressure acting as a resistance at the entry should be conquered first and then spontaneous capillary pressure acting as a driving force can trigger further a series of spontaneous capillary events, e.g., touch and overlap, etc. [37]. Thus, we propose another pressure ratio, $R_c$, to describe the competition between capillary resistance and spontaneous driving, i.e.,

$$R_c = \frac{P_c^{max}}{P_c^{min}}, \tag{9}$$

where $P_c^{max}$ and $P_c^{min}$ are the maximum and minimum capillary pressure in Eq. (7), respectively.

Combining $R_T$ and $R_c$, we obtain a complete description of effects on capillary-dominated fluid-fluid displacement in hierarchical porous media, and a so-called "hierarchical number" is expressed as

$$\text{Hi}^c = R_T \cdot R_c. \tag{10}$$

Compared to the hierarchical number proposed in our previous work for viscous fingering [30], besides some similarities in constructing both hierarchical numbers, $\text{Hi}^c$ here shows certain variation on terms. Specially, the driving force becomes different since the capillary number is low enough here, i.e., the fluid-fluid displacement in porous media is a spontaneous process driven by the capillary suction in both 1st- and 2nd-order porous structures. Whereas, for viscous fingering with high capillary number, the flow in 1st-order porous structure is mainly forced by the external pressure while capillary suction in 2nd-order porous structures. For both viscous and capillary fingering cases, with a larger hierarchical number more connections of invading fluid can be established during displacement so that instability induced by the disorder of obstacle arrangement in the capillary-dominated situation, or by viscosity variation in the viscous-dominated situation, can be alleviated.

## 3.4 Correlation analysis

To further quantify the relationship between the proposed hierarchical number and displacement patterns, the adopted indexes in Section 3.2 as function of $Hi^c$ are demonstrated in Figure 4. The crossover boundary shown in Figure 2 corresponds to $Hi^c = 6.5 \times 10^8$. overall, as shown in Figure 4, all indexes present a monotonous variation with $Hi^c$. Specifically, the growth rate of interface length, $k_{int}$, is inversely proportional to $Hi^c$; the relationship between $Hi^c$ and stable fractal dimension $\widetilde{D}_f$ or 2$^{nd}$-order saturation rise rate, $k_{sat}$, is a S-shaped curve, which marks the upper and lower bounds as two asymptotes corresponding to the extreme values in compact and fingering modes, respectively. Compared to the monotonically decreasing curve of $k_{int}$, the upper asymptote in S-shape curves suggests that $\widetilde{D}_f$ and $k_{sat}$ cannot reflect the slight difference induced by local instability like fluid trapping presented in the strong imbibition regime.

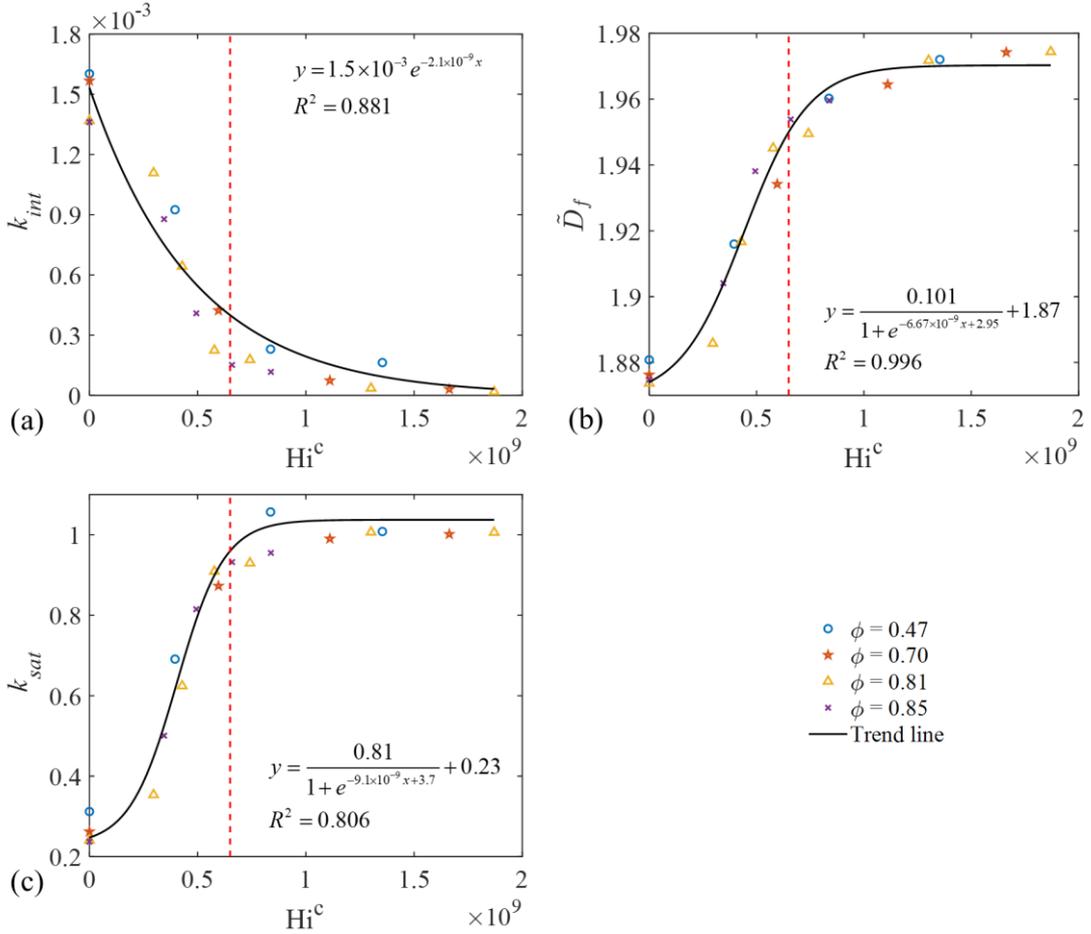

Figure 4. Flow pattern indexes vs. hierarchical number, i.e., (a) growth rate of interface length, $k_{int}$; (b) fractal dimension, $\widetilde{D}_f$; (c) rise rate of 2nd-order saturation with 1st-order saturation, $k_{sat}$. The crossover boundary is marked by the dash line, as $\text{Hi}^c = 6.5 \times 10^8$.

## 4. Conclusion

This work focuses on the capillary-dominated displacement pattern in hierarchical porous media. Besides the effects of the wettability and geometrical disorder, we mainly investigated the impacts of hierarchical geometry on the transition of displacement modes through a series of pore-scale simulations. Based on the simulation results, the following conclusions can be reached:

(1) The displacement pattern can be classified into the compact mode and fingering mode and a phase diagram, demonstrating the mode transition under combined conditions of wettability and geometry feature, suggests that weakening the capillary effects by increasing the 2$^{nd}$-order porosity can enhance the displacement stability;

(2) The mechanism of fingering suppressing is clarified based on the analysis regarding the displacement-related indexes and specifically the wetted 2$^{nd}$-order porous space may bring in extra connections within invading fluid leading to a more uniform distribution of capillary pressure, so that the interface advances in a stable and compact mode;

(3) Finally, a dimensionless number ($\text{Hi}^c$) is proposed here to quantitatively estimate the involvement of 2$^{nd}$-order pore space during the displacement by considering capillary-induced resistance and driving force together.

Complimented with the work on viscous fingering, we can provide a complete picture on the fluid-fluid displacement control, and specifically if fingering is expected to be suppressed in a capillary-dominated situation, the capillary effects should be weakened, e.g., by loosening the packing in 2$^{nd}$-order space; inversely, if in a viscous-dominated situation, the capillary effects should be enhanced, e.g., densifying the packing in 2$^{nd}$-order space. Since the hierarchical porous structure is commonly encountered in nature and applications [38-40], the mechanism of displacement transition may contribute to linking the pore-scale observation with the multiphase flow at field scale. In addition, this study is also potential to benefit the design of microfluidic devices [41-43] to have tailored flow patterns.